\documentclass[aps,prl,reprint]{revtex4-1}

\usepackage{graphicx}

\begin{document}

\title{Dipolar coupling between nanopillar spin valves and magnetic quantum cellular automata arrays}

\author{Madalina \surname{Colci}}
\email{colci@illinoisalumni.org}
\author{Mark B. \surname{Johnson}}
\affiliation{Naval Research Laboratory, Washington, DC 20375, USA}

\date{\today}

\begin{abstract}
We experimentally demonstrate magnetostatic coupling between a nanopillar pseudo spin valve structure and a linear array of dipole coupled Permalloy nanomagnets. Using magnetic force microscopy, we study the interaction between the spin valve and the first element of the array, and present evidence that the nanomagnet couples with the hard layer of the spin valve for two spin valves with distinctly different composition. Our study includes a statistical analysis of antiferromagnetic order within the linear array, and provides insight into the range of behavior that these arrays can display. These results bear directly on the design of magnetic quantum cellular automata (MQCA) logic devices, showing that multilayer devices can couple to simple nanomagnets. Redesigning the hard layer of the magnetoresistive devices would make them operational as an electronic input that will allow integration of MQCA networks in complex electronic circuitry.
 \end{abstract}

\maketitle

\section{I. Introduction}

A strong candidate for a low power alternative to complementary metal oxide semiconductor (CMOS) information processing is a magnetic quantum cellular automata (MQCA) logic cell based on magnetostatically-coupled nanomagnets \cite{Cowburn00, Imre06}. In this approach, binary logic states (0, 1) are associated with the bistable magnetization states of a patterned ferromagnetic element that has a uniaxial anisotropy axis. The operating power is very low because there is no transport of electric charge. MQCA prototype cells have already demonstrated primitive Boolean operations \cite{Imre06}. However, these basic MQCA cells are operated in isolation, with external magnetic fields used to set the initial state of the driver nanomagnet, and logic outcome observed by magnetic force microscopy (MFM). In order to be computationally useful, they must be interfaced with electronic circuitry for integrated input and output (I/O).
	
\begin{figure}
\includegraphics[width=1.0\linewidth]{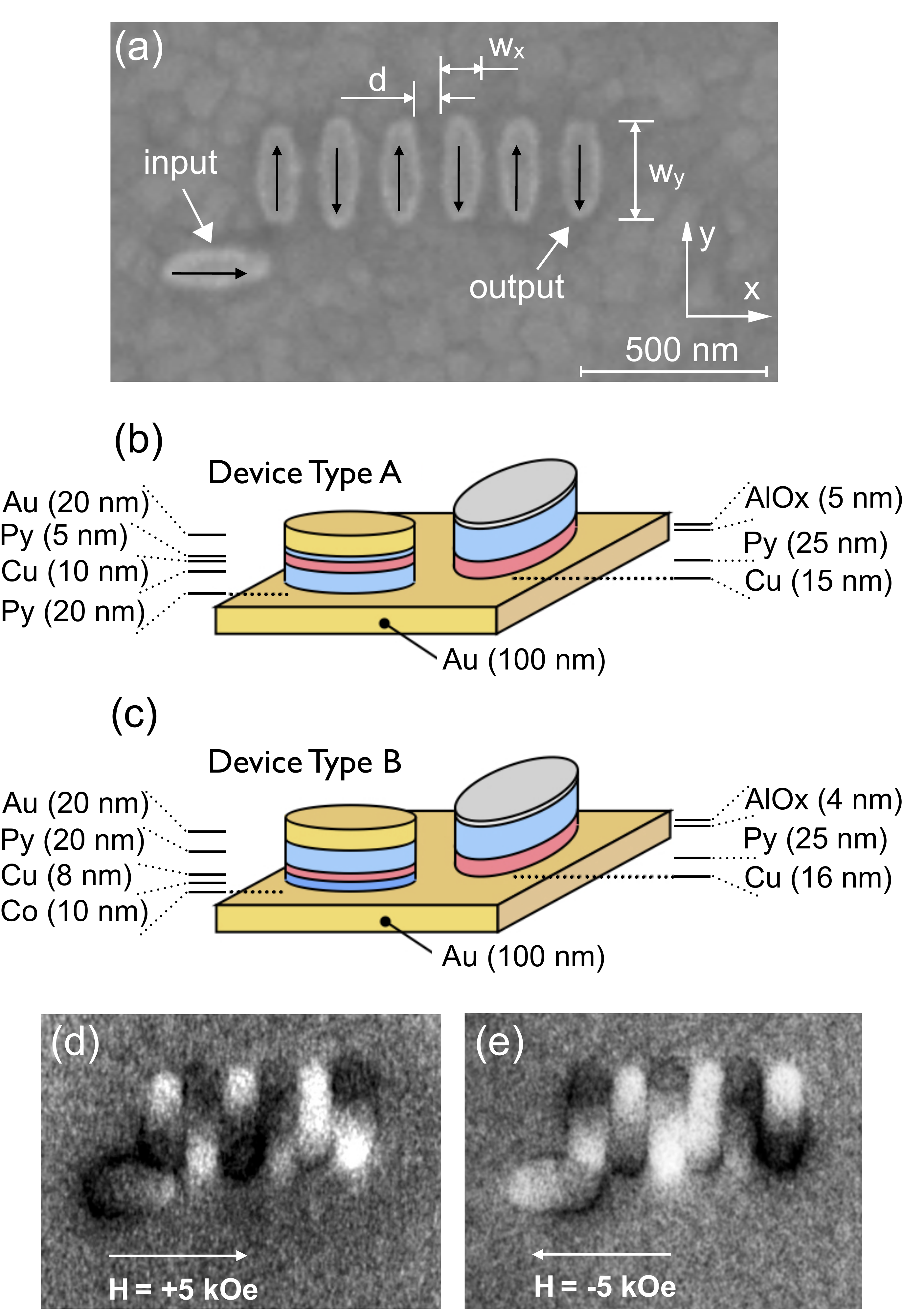}
\caption{ (Color online) (a) Scanning electron micrograph of a typical MQCA cell comprising an input element and a nanomagnet linear array. The arrows superimposed on the elements schematically show antiferromagnetic order in the chain. (b), (c) Layer structures of (b) Type A and (c) Type B cells. (d), (e) MFM images of one MQCA cell for two opposite magnetization states of the input spin valve. The light and dark contrast areas correspond to the two magnetic poles of single domain magnets. The arrows indicate the external magnetic field direction applied prior to imaging.}
\label{fig1}
\end{figure} 
	
	A promising approach for integrated I/O incorporates magnetoresistive (MR) elements such as current-perpendicular-to-the-plane (CPP) spin valves (SVs) or magnetic tunnel junctions (MTJs). These elements could be introduced at appropriate positions in the MQCA array, and electrical contact could be made with conductive nanowires. Output could be achieved by sensing the resistive state of the MR element, which correlates with the relative magnetization orientation of the magnetic layers. Input could be performed by switching the magnetization orientation of the free layer using spin-torque transfer (STT) currents, thus minimizing the risk of influencing other magnetic elements on the chip. This approach requires demonstration of STT switching and MR sensing of an appropriately sized and shaped MR device. This has been accomplished for isolated CPP spin valve nanopillars \cite{Katine00, Albert00, Grollier01, Wegrowe02, Sun02}. Also, an MTJ-based input has been studied using STT programming in a one-dimensional array made entirely of MTJs \cite{Lyle1, Lyle2}. In this case, MTJs were fabricated for use as the main computation devices and not just the I/O interface device. 
	
	The fundamental factor that would enable an integrated I/O as discussed above is the magnetostatic coupling between the MR device and the nanomagnets in the MQCA array. To date, this has not been experimentally demonstrated. The focus of our study is the design and implementation of a simple MQCA cell consisting of a pseudo spin valve (PSV) and an adjacent linear array of nanomagnets to show proof of the concept that a multilayer magnetic element can be used as the input stage of MQCA. We experimentally demonstrate coupling between the SV element and the array using MFM measurements. We analyze the efficiency of PSV coupling to the array, and report on the switching characteristics of the array with a discussion of the information transfer along the array and a limited statistical analysis of switching success rates.
		
\section{II. Experiment}
	
		Of the several MQCA circuits that have been proposed \cite{Porod1}, a relatively simple sub-circuit is the linear array of antiferromagnetically coupled nanomagnets with a single input (or output) element that is magnetostatically coupled to the first element of the array \cite{Porod2}. Shown in Fig. \ref{fig1}(a), each nanomagnet is an ellipse with an aspect ratio of 2.5. In our cells, the nanomagnets' width is $w_x = 60$ nm, and the separation $d$ between ellipses is a variable, 15 nm $\leq d \leq$ 30 nm. The nanomagnet array is arranged along the $x$-axis, with the long axis of each ellipse oriented along the $y$-axis. For the physical arrangement of nanomagnets with long edges in parallel, as shown in Fig. 1(a), and for a chosen magnetic material, thickness and separation $d$, magnetostatic coupling between the neighbors determines the low energy stable state of the one dimensional Ising array to have antiferromagnetic (AF) order. This state can have either one of the two possible configurations of alternating magnetic dipoles.
			
	We fabricate a PSV as the input element at a distance of approximately $d$ from the first nanomagnet. The long axis of the input element is oriented along the $x$-axis. If the input element is sufficiently close to the array, then its fringe magnetic field determines the magnetization direction of the first nanomagnet in the chain. This effectively sets the logic state of the array because the first nanomagnet, through magnetostatic coupling with the nearest neighbor, determines which of the two possible AF configurations is adopted by the array. The coupling inside the array represents the information propagation stage of MQCA. This aspect of MQCA has already been studied extensively; in our experiment we focus on the coupling between the multilayer device and the first nanomagnet in the chain.

	As seen in the cross section views of Figures \ref{fig1}(b) and \ref{fig1}(c), the multilayer structure of the magnetoresistive element differs from that of the nanomagnet. Therefore, a nanomagnet adjacent to the multilayer device may couple to either the hard (pinned) or soft (free) layer of the MR element, or coupling may fail. Integrated input/output of the MQCA cell requires coupling to the free layer.
		
	The first step of sample fabrication is the formation of a patterned gold film, 100 nm thick, which serves as a bottom electrode for the pseudo spin valve [Figs. \ref{fig1}(b) and \ref{fig1}(c)]. In the second step the magnetoresistive input element is fabricated in an additive approach.  An ellipse is defined using electron-beam lithography and PMMA resist. The multilayer materials stack is deposited by electron-beam evaporation in a base pressure of $7 \times 10^{-8}$ torr. Finally, the resist is stripped leaving the nanopillar spin valve standing on the bottom electrode. 
	
	Two different pseudo spin valves are used. In Device Type A [Fig. \ref{fig1}(b)] the multilayer is a standard PSV nanopillar composed of Py(20)/Cu(10)/Py(5)/Au(20); Py denotes Permalloy (Ni$_{80}$Fe$_{20}$), and the thickness is in nm. The top Py layer has a smaller coercivity than the bottom layer and acts as the free layer. In Device Type B [Fig. \ref{fig1}(c)], composed of Co(10)/Cu(8)/Py(20)/Au(20), the Co layer has larger intrinsic coercivity and acts as the pinned layer. The thickness of the top Py film, the free layer, was increased to promote coupling between the free layer and the nanomagnet array. The Au cap layer passivates the spin valve and facilitates electrical contact to a top electrode. 
	
	In the third fabrication step the nanomagnet array is formed using similar processes of e-beam lithography, e-beam deposition, and liftoff. The Py nanomagnets are placed on Cu pedestals with thicknesses chosen such that the nanomagnet is positioned next to the top layer of the PSV. Therefore, the resulting Cu(16)/Py(25)/Al(4) structure facilitates magnetostatic coupling with the free layer of the PSV. The aluminum cap is oxidized and forms an electrically insulating passivation layer. If the top electrode of the PSV were to overlap one or more of the Py nanomagnets, the aluminum oxide cap would inhibit conduction through the nanomagnets and permit electrical contact only through the spin valve. As seen in Figure \ref{fig1}(a), precise lithographic alignment along $\hat{y}$ (50 nm or better) of the PSV and the nanomagnets is achieved.

	\section{III. Results}
	
	\begin{figure}
\includegraphics[width=1\linewidth]{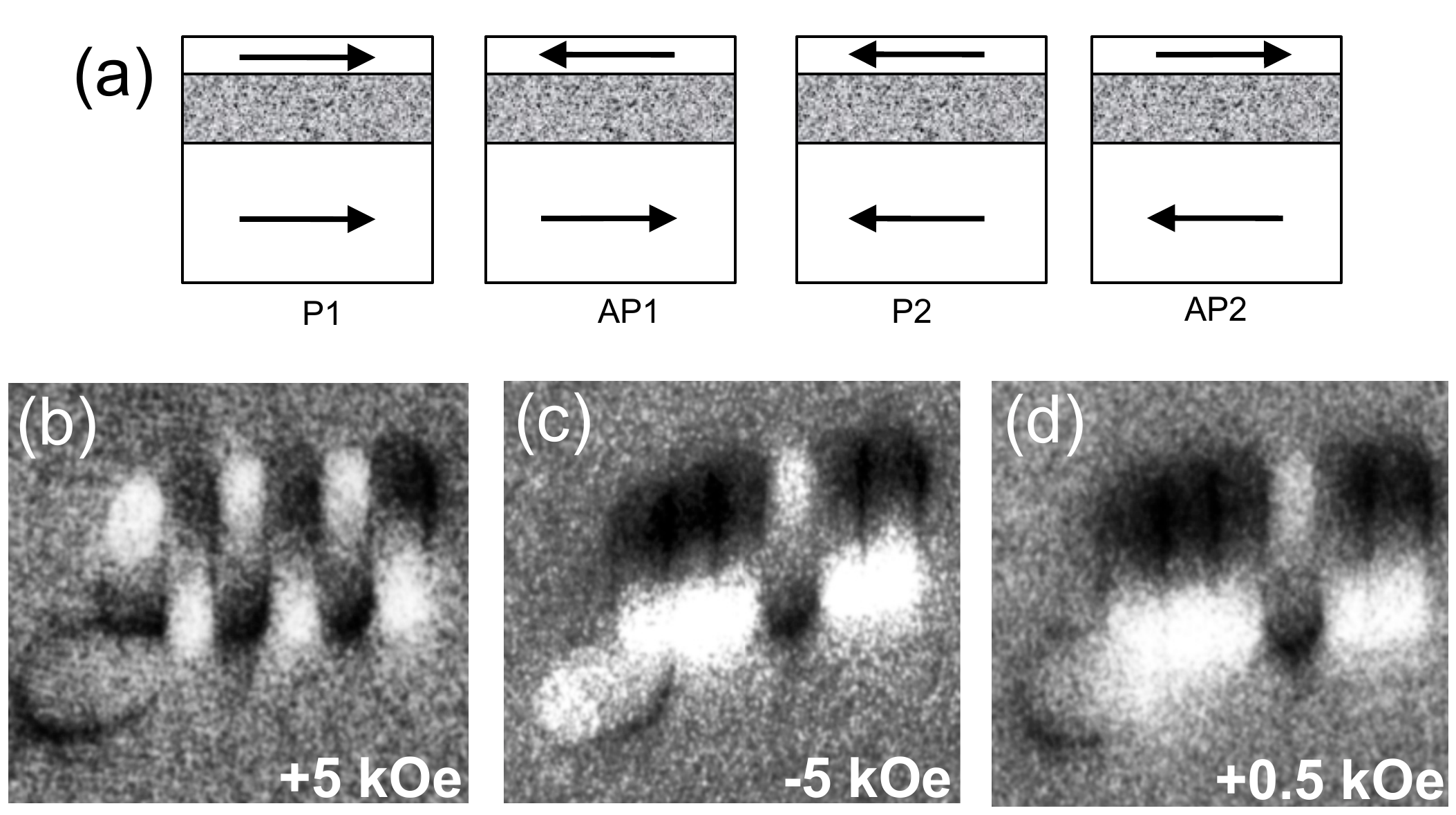}
\caption{(a) Schematic cross section of a type A spin valve showing the four possible magnetic states. The arrows illustrate the magnetization orientations in the bottom, Py(20 nm), and top, Py(5 nm), layers. (b)-(d) MFM images of a type A MQCA cell in different magnetic states determined by the magnetization state of the input spin valve.}
\label{fig2}
\end{figure} 
	
	We perform MFM imaging on the MQCA cells with no top electrode in place. First, both magnetic layers in the PSV are saturated to have parallel orientation by applying a large external magnetic field $H = \pm$ 5 kOe along the $x$-axis. The remanent magnetization state of the cell ($H = 0$) is then imaged. We use low-moment MFM tips in the tapping mode with a lift scan height of 20 nm to record the amplitude and phase-shift of the cantilever oscillation. The bipolar contrast in the MFM images [refer to Figs. \ref{fig1}(d) and 1(e)] is associated with the two magnetic poles of a single domain magnet when the magnetization orients along the long axis of the ellipse. As expected due to shape-induced magnetic anisotropy, the nanomagnets in the array are single domain. Figures \ref{fig1}(d) and  \ref{fig1}(e) show MFM images of a Type B MQCA cell after the spin valve is magnetized in large positive and negative fields, respectively. In Figure \ref{fig1}(d), the PSV magnetization lies along $+\hat{x}$, couples to that of the first nanomagnet, and antiferromagnetic order is induced in the array. In Figure \ref{fig1}(e), the magnetization of the PSV reverses to lie along $-\hat{x}$. Magnetostatic coupling causes the reversal of the magnetization orientation of the first nanomagnet in the array, and the antiferromagnetic order of the linear array is successfully reversed. This experiment demonstrates that a multilayer magnetoresistive element can provide sufficient field strength to switch a nanomagnet in the chain and therefore can be used to set the state of an ordinary MQCA nanomagnet array.
		
	The results in Figs. \ref{fig1}(d) and  \ref{fig1}(e) do not provide information about which layer of the spin valve couples to the first nanomagnet in the array. We derive this detail by systematically varying the magnitude of $H$ and then recording MFM images at remanence. Figure 2(a) schematically presents the four possible magnetic states in a type A spin valve. Figures \ref{fig2}(b)-(d) show three MFM images of a type A cell acquired after applying $+5$ kOe, $-5$ kOe, and $+0.5$ kOe, in this order, along the $x$-axis. After each field sweep we image the remanent state. In Figure \ref{fig2}(b), the PSV is in a remanent state P1 and antiferromagnetic order in the nanomagnet array has been induced. In Figure \ref{fig2}(c), the magnetization orientation of the Py layers is reversed and the PSV is in a remanent state P2. Magnetostatic coupling properly sets the magnetization orientation of the first nanomagnet, but the long-range antiferromagnetic order in the array is lost; we give possible reasons for this in the discussion section. After applying a field $H = +0.5$ kOe [Fig. \ref{fig2}(d)], the magnetization configuration of the PSV has changed but the orientation of the first nanomagnet has not changed. This suggests that the top Py layer of the spin valve has reversed, but the nanomagnet is coupled to the magnetization of the bottom Py layer and its orientation is unchanged. This experiment demonstrates that while a Type A PSV can be used to set the state of a MQCA nanomagnet array, it is unsuitable for use as an input-interfacing device since coupling is not achieved via the free layer. 
		
	We note that in Figures \ref{fig2}(b)-(d) the spin valve does not show a strong magnetic contrast as seen in the Py elements. This is expected since the signal detected by a magnetic force microscope is proportional to the second derivative of the stray field of the magnet. Therefore, the fringe field from the thin top layer of the PSV produces a significantly smaller signal than that from the thick nanomagnets.

\begin{figure}
\includegraphics[width=1\linewidth]{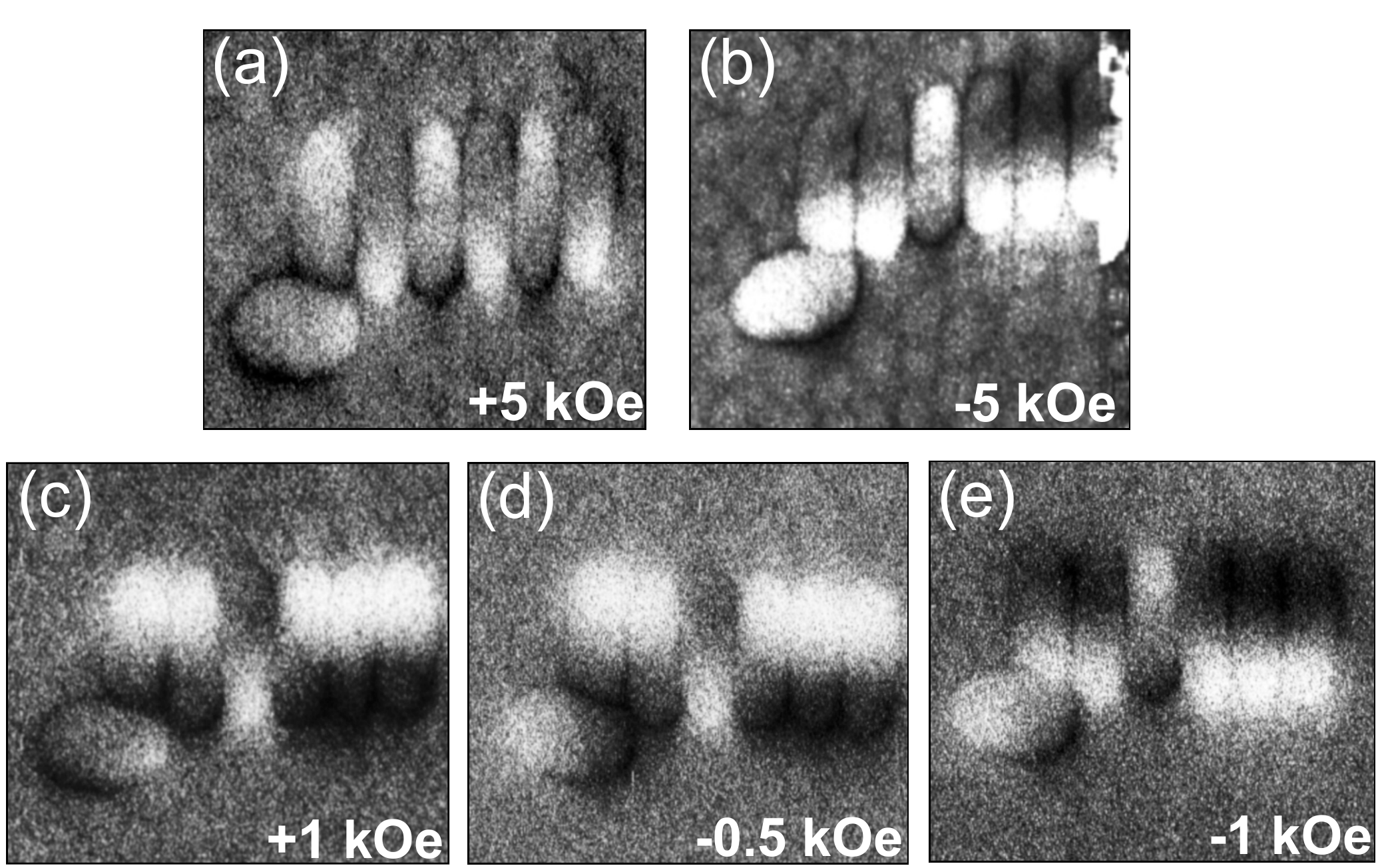}
\caption{(a)-(e) MFM images of a type B MQCA cell in different magnetic states determined by the magnetization state of the input spin valve, as discussed in the text.}
\label{fig3}
\end{figure}

	A Type B spin valve in which the thickness of the free layer is comparable to that of the nanomagnets may be more likely to achieve the desired coupling between the free layer and the first nanomagnet. We investigate this next using the MFM images shown in Figure \ref{fig3}. The parallel and antiparallel magnetic states of a Type B PSV are the same as in Figure \ref{fig2}(a), even though the layers have different thicknesses compared with Device Type A. This is due to the fact that the bottom Co film has a higher coercivity than the top Py layer, thus acting as the hard layer. The images in Figure \ref{fig3}(a)-(e) are acquired after applying external fields $H = +5, -5, +1, -0.5$ and $-1$ kOe, in this order. In Figure \ref{fig3}(a) the PSV is in a remanent state P1 and the antiferromagnetic order in the nanomagnet array has been induced. This demonstrates that a PSV of Type B also can be used to initialize the state of a MQCA nanomagnet array. In Figure \ref{fig3}(b) the spin valve is in state P2 and magnetostatic coupling properly sets the magnetization orientation of the first nanomagnet, but complete antiferromagnetic ordering in the array is not achieved. We examine this behavior in the discussion section. Next, we sweep the field to $+1$ kOe and then image the remanent configuration. We see in Figure \ref{fig3}(c) that the magnetization orientation of the PSV has reversed as well as that of the first nanomagnet, indicating successful magnetostatic coupling. We further note that the orientation of each nanomagnet in the array is reversed, even in the presence of mixed antiferromagnetic and ferromagnetic ordering.

	In order to identify which layer of the PSV is responsible for the switching of the array we apply smaller external magnetic fields to find the regime where only one of the magnetic layers of the PSV has reversed its magnetization. Following the above field sweep to $+1$ kOe we reverse the field polarity and subsequently apply $H = -0.2, -0.5$ and $-1$ kOe. We record remanent state images after each field. We see no change after applying $-0.2$ kOe (image not shown here), indicating that this field is smaller than the coercive field value of both magnetic layers. Upon increasing the magnitude of the field further to $-0.5$ kOe, the magnetic state of the spin valve is reversed, while the magnetization in the array has not changed [Fig. \ref{fig3}(d)]. Since the top layer of these structures is the soft layer, the spin valve is now in the AP1 state. This image suggests that the nanomagnet array remains coupled to the bottom Co layer of the PSV. In the final step, shown in Fig. \ref{fig3}(e), the MFM image of the PSV is the same as that in Fig. \ref{fig3}(d), but the magnetization orientation of each element of the array has reversed. We interpret this as implying that the orientation of the hard layer reversed, causing the orientations of the nanomagnets, which are coupled to it, to also reverse.
		
	\section{IV. Discussion of Results}
	
The dipole-dipole interaction between the input magnet and the first nanomagnet causes the south pole of one magnet to align with the north pole of the adjacent magnet. The resulting magnetization pattern leads to alternating black and white contrast in the MFM images. Based on this contrast we find that all but one of the images in our study show correct dipolar coupling between the spin valve and the first element of the array, which represents a 97$\%$ coupling success rate. However, since this coupling is achieved via the hard layer, the present multilayer structure is not optimal for operation as an input. A modified layer structure that creates small or no fringe fields from the fixed layer should be employed in future studies.

\begin{figure}
\includegraphics[width=0.63\linewidth]{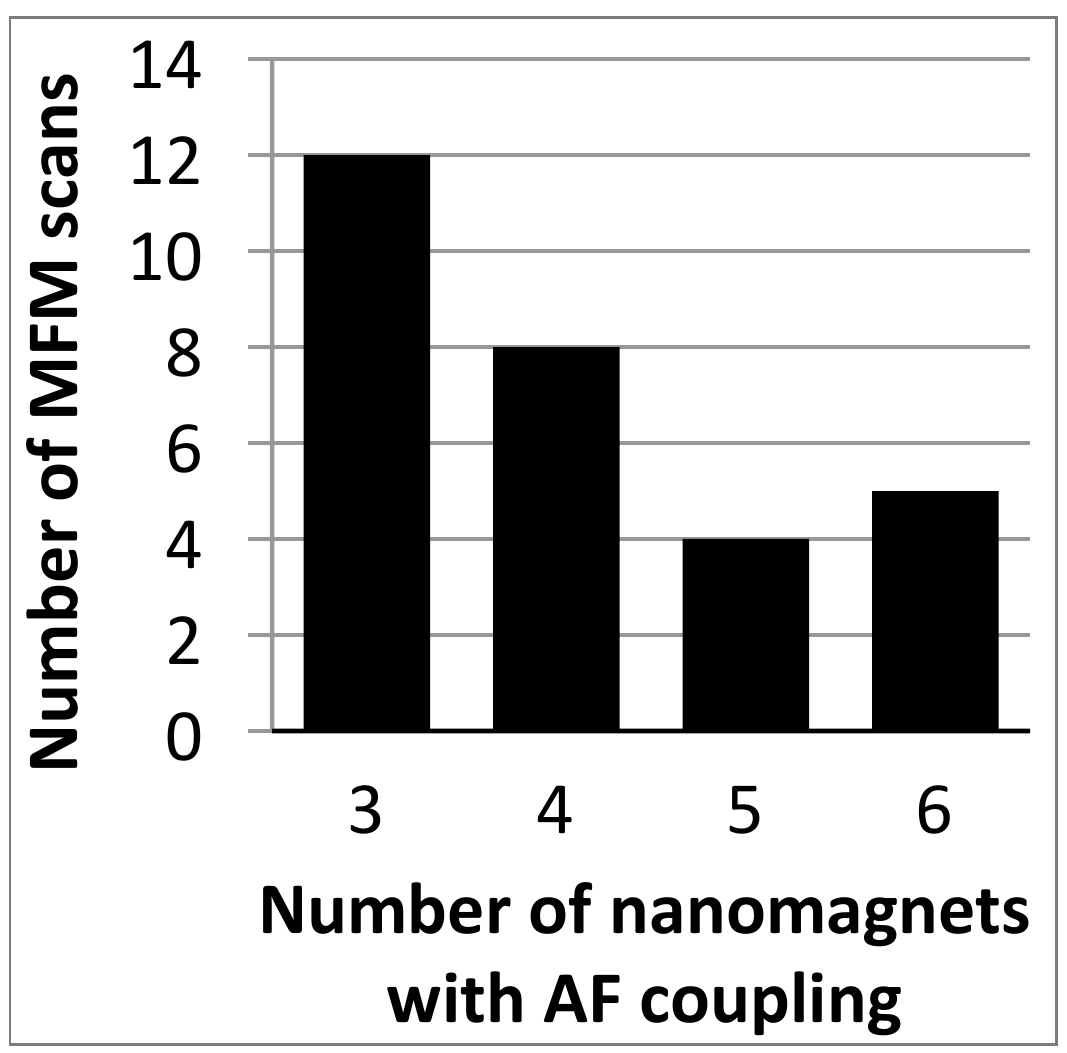}
\caption{Histogram of the coupling behavior in our nanomagnet arrays. One scan with zero AF-coupled nanomagnets is not included in this histogram. }
\label{fig4}
\end{figure}

	Figure \ref{fig4} summarizes the coupling behavior in the arrays derived from 30 MFM scans. Full AF ordering between all six nanomagnets is achieved in 17$\%$ of the scans, while 40$\%$ of them show coupling between only three nanomagnets. This result is consistent with previous studies of long chains of nanomagnets which find a typical correlation length of $4-7$ magnets \cite{Cowburn02, Imre03}, although long range order in 20 nanomagnets has been observed \cite{Imre34}. 
	
\begin{figure}
\includegraphics[width=0.72\linewidth]{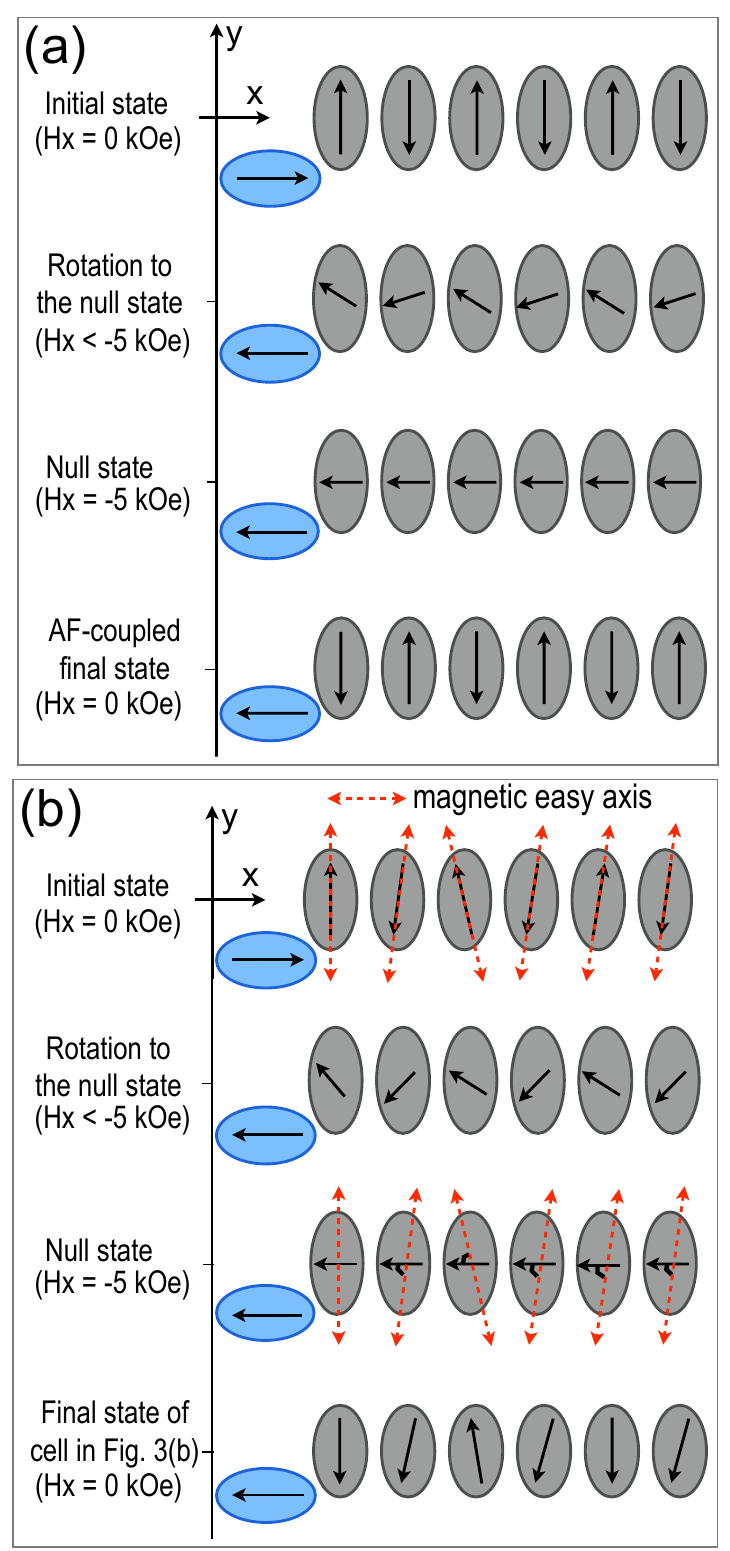}
\caption{(Color online) (a) Schematic of a nanomagnet chain undergoing a demagnetization procedure by an external magnetic field $H$ applied along the hard axis (i.e., $x$-axis) of the nanomagnets. The final magnetization state after the removal of the external field is the result of the dipole coupling between nearest neighbors. (b) Schematic of a nanomagnet chain in which the nanomagnets have a misaligned magnetic easy axis, depicted by the dashed, double-arrow lines. The rotation from the null state to the final state occurs via the smallest angle of rotation, indicated in the null state diagram. The final magnetization state corresponds to that in Fig. 3(b).}
\label{fig5}
\end{figure}

	The most notable behavior in our cells is the asymmetric switching of the array: the perfect AF coupling after application of $+5$ kOe is lost after the field is swept to $-5$ kOe. This type of corruption of data transfer was previously reported in long chains of Py nanomagnets and studied as a function of the demagnetizing process for various magnet shapes \cite{Imre34}. Indeed, the magnetic field used to set the magnetization of the input spin valve is one of the most important parameters of the system operation since it also acts as a demagnetizing field for the chain of nanomagnets. We illustrate the demagnetization process in the schematic in Fig. \ref{fig5}(a). Starting with the nanomagnets in an AF-coupled state, we apply the external field along their hard axis.  The magnetization of the nanomagnets rotates toward the hard axis, and is finally forced to lie along this axis. This magnetic state of the array is called the ``null'' logic state. Next, when slowly removing the magnetic field, the stray field from the input element tilts the magnetization of the first nanomagnet out of its null state. As its dipole field rotates, this nanomagnet tips the second nanomagnet off its hard axis toward the easy magnetization axis. Every nanomagnet in the chain takes its turn to rotate in the direction dictated by the stray field from the nearest neighbor. If there are no defects, this chain will align antiferromagnetically once the external magnetic field is removed (AF-coupled final state in Fig. \ref{fig5}(a)). 
	
	The success of the information transfer described above relies on the stability of the magnetization along the hard-axis until the signal reaches each nanomagnet. Since the null state is metastable, it can be relatively easily influenced by various factors. For example, the magnetic field applied along the hard axis modulates the energy barrier between the up and down magnetization states. However, above a critical value of the field, this energy barrier becomes lower than the thermal energy, thus rendering the magnetization dynamics susceptible to thermal fluctuations \cite{Brown63}. The fluctuations may cause the magnetization to rotate into an easy axis direction before the stray field from a neighboring nanomagnet can tip it in the correct, dipolar-coupled direction. These spontaneous transitions of the magnetization can therefore lead to ferromagnetically-aligned pairs in an array that should be AF-coupled.

	Another way in which errors can be introduced in the array is by having a small angular deviation of the easy magnetization axis of a nanomagnet from the geometric major axis of the ellipse. Simulations of nanomagnet chains in which each nanomagnet has a different  small angular misalignment have shown that easy axis alignment is a cause of errors in uniaxial nanomagnets, with high error probability for even 2-3 degrees of deviation from the  ellipse's major axis \cite{Carlton12}. These simulations also predict that the error's positioning is repeatable upon multiple magnetic field cycles, which is what we observe in our experiment. 
		
	Figure \ref{fig5}(b) schematically shows how the misalignment discussed above can alter the information transfer, and result in the magnetization configuration of the cell in Fig. \ref{fig3}(b). We depict the magnetic easy axis by the dashed line with a double-ended arrow, and place it at a small angle with respect to the major axis of the elliptical nanomagnet. The rotation to the null state proceeds just as in Fig. \ref{fig5}(a). For the null state we draw the smallest angle between the hard axis and the magnetic easy axis. In the next step, when the demagnetizing field is reduced, the magnetization will rotate from the hard axis onto the easy axis direction determined by the smallest angle of rotation because this path minimizes the energy dissipation associated with magnetization reversal.

	One factor that can cause easy axis misalignment is the microstructural variations of the Py layer caused by the surface roughness of the Cu pedestals on which the nanomagnets are built. Such roughness can lead to interface granularities that have different magnetic moments. The exchange interaction between neighboring grains can then cause the magnetization direction to deviate from the major axis. Another factor that can destabilize the easy axis orientation is lithographic irregularities, such as shape nonuniformities and edge roughness. These are unavoidable variations when fabricating nanostructures. Edge roughness of even a few nanometers has been shown to lead to a second anisotropy term which favors the magnetization to align with the shorter edge of an elongated nanostructure \cite{Cowburn002}. This term strongly competes with the demagnetizing field due to shape anisotropy, and can result in a misaligned easy axis.

\begin{figure}
\includegraphics[width=1\linewidth]{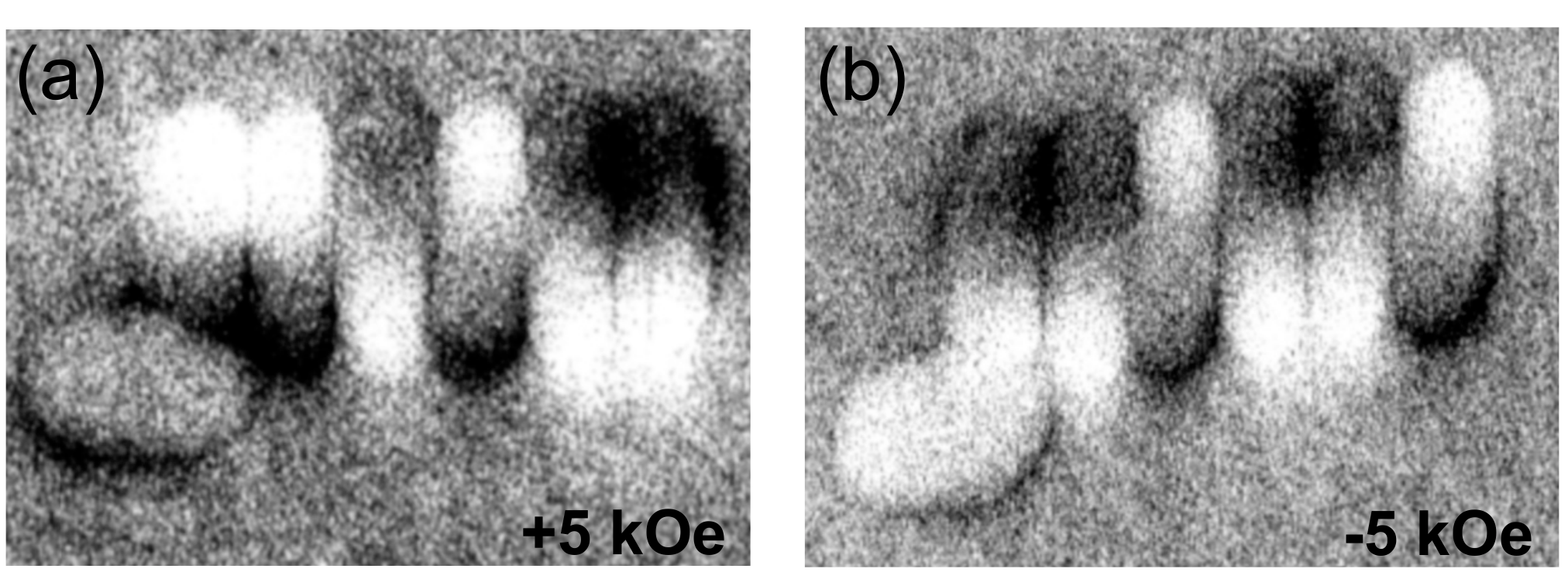}
\caption{(a), (b) MFM images of one MQCA cell in two different magnetization states. The last element of the array has the correct magnetization state expected from correct coupling between the input spin valve and the array, even though AF coupling is not established along the entire chain.}
\label{fig6}
\end{figure}

	In addition to the asymmetrical switching explained above, we observe another type of behavior in some of our arrays. Figures \ref{fig6}(a) and \ref{fig6}(b) show MFM images of a type A cell in which the input PSV is programmed in the two parallel states P1 and P2, respectively. We see that the array is properly initialized but that the first two nanomagnets are ferromagnetically coupled to each other, and switch together when the PSV reverses its magnetization. While the types of fabrication irregularities discussed above can explain this behavior, there is also another possibility: a displacement along $\hat{x}$ of the array with respect to the spin valve. Ideally, the leading edge of the first nanomagnet should align with the tip of the PSV, as seen in the SEM image of Figure \ref{fig1}(a). If a significant displacement is present, the influence of the PSV fringe field may extend to the second element of the array. We also note that the last element in Fig. \ref{fig6}(b) has reversed its orientation in a way consistent with correct information transfer along the chain as set by the input. However, not all elements are coupled antiferromagnetically. This experiment shows that care must be taken when interpreting magnetoresistance data of a readout MR device: even though the output of the array may have the expected magnetization state, it does not imply perfect AF coupling along the entire chain. MFM should be used in this early stage of developing electronic I/O for MQCA to confirm error-free coupling of the nanomagnets.
	
\section{V. CONCLUSIONS AND FUTURE DIRECTIONS}
	In summary, we present an MFM study of a linear array of nanomagnets in close proximity to a nanopillar spin valve. We demonstrate antiferromagnetic coupling between the elements in the array, and between the spin valve and the adjacent nanomagnet. However, we observe that the dipole interaction between the hard layer of the PSV and the first nanomagnet dominates the coupling. Proper operation of the PSV as an input requires the free layer to couple to the nanomagnet array. This may be achieved in future experiments if the fixed layer is a synthetic antiferromagnetic structure, thus creating no fringe fields when it is perfectly balanced. 
	
	Our results also show errors in data propagation along the MQCA chain. The failure of nominally identical nanomagnets to couple is generally understood in terms of differences in microstructure, edge roughness, and irregularities in shape present in lithographically defined nanomagnets. Increasing the coupling between the neighboring nanomagnets, either by decreasing their separation or by increasing their volume, could make the chains more stable against errors. Another approach that has already been proposed is to engineer biaxial anisotropy nanomagnets, thus enhancing the stability of the magnetization along the hard axis against thermal fluctuations \cite{Carlton08}. All these improvements will lead to MQCA networks adequate for practical applications.	
	
	\section{ACKNOWLEDGMENTS}
	
	This work was supported by the Office of Naval Research Grant No. N0001409WX30420. The experiments were carried out at the Naval Research Laboratory. We thank M. Miller for help in starting the MFM measurements, and the NRL Nanoscience Institute for use of equipment and instruments.

\end{document}